# Evolution of Continuum from Elastic Deformation to Flow

Xiao Jianhua

Natural Science Foundation Research Group, Shanghai Jiaotong University

**Abstract:** Traditionally, the deformation of continuum is divided into elastic, plastic, and flow. For a large deformation with cracking, they are combined together. So, for complicated deformation, a formulation to express the evolution of deformation from elastic to flow will help to understand the intrinsic relation among the related parameters which relate the deformation with a stress field. To this purpose, Eringen's polar decomposition and Trusedell's polar decomposition are formulated by explicit formulation of displacement field, based on Chen's additive decomposition of deformation gradient. Then the strain introduced by the multiplicative decomposition and the strain introduced by the additive decomposition are formulated explicitly with displacement gradient. This formulation clears the intrinsic contents of strains defined by taking the Eringen's polar decomposition and Trusedell's polar decomposition. After that, it shows that the plastic deformation can be expressed as the irreversible local average rotation. For initial isotropic simple elastic material, the path-dependent feature of classical plasticity theory is naturally expressed in Chen's strain definition. It is founded that for initially isotropic material the motion equations require a non-symmetric stress for dynamic deformation and a symmetric stress for static deformation. This controversy between dynamic deformation and static deformation can be used to explain the cracking or buckling of solid continuum. Finally, the research shows that the flow motion of continuum can be expressed by the same formulation system. So, it forms an evolution theory from elastic deformation to flow of continuum.

## 1 Introduction

Traditionally, the deformation of continuum is divided into elastic, plastic, and flow. Consequently, the mechanics of continuum establishes several branches. Along with the development of high-tech industry, complex deformation becomes the interested topic. For complicated deformation, a formulation to express the evolution of deformation from elastic to flow will help to understand the intrinsic relation among the related parameters which relate the deformation with a stress field. This will leads to non-linear theory of mechanics[1].

For large deformation, Eringen's polar decomposition and Trusedell's polar decomposition are widely used to formulate strain definition[2-6]. However, their explicit formulations by displacement field are not available. Moreover, the geometric interpretation of orthogonal rotation tensor is too mathematic to be phenomenal. This disadvantage dragged the development of continuum mechanics for a continuous deformation from infinitesimal to finite flow. Recently, this problem is attacked by different ways.

Based on Chen's additive decomposition of deformation gradient[7-11], the strain introduced by the multiplicative decomposition and the strain introduced by the additive decomposition are formulated explicitly with displacement gradient. This formulation clears the intrinsic contents of strains defined by taking the Eringen's polar decomposition and Trusedell's polar decomposition.

Introducing Chen's strain definition, it is shown that the plastic deformation can be expressed as the irreversible local average rotation. For initial isotropic simple elastic material, the path-dependent feature of classical plasticity theory is naturally expressed in Chen's strain definition. Therefore, constitutive equations (including elastic, plastic, rheology, and flow) can be nationalized with the new understanding.

It is founded that for initially isotropic material the motion equations require a non-symmetric stress for dynamic deformation and a symmetric stress for static deformation. This controversy between dynamic deformation and static deformation can be used to explain the cracking or buckling of solid continuum. Finally, the research shows that the flow motion of continuum can be expressed by the same formulation system. So, it forms an evolution theory from elastic deformation to flow of continuum.

In a typical experiment, for a complete finite deformation process of first loading, the named stress and the named strain are related with a non-linear relation. However, the striking feature of the unloading process is that they show an invariant elastic relation referring to the plastic strain, which is determined by the maximum deformation the material suffered. The reloading process firstly follows the elastic path and then, at the point of maximum deformation, follows the original path of finite deformation.

In classical plasticity theory, this path-dependent stress-strain relationship is expressed by introducing plasticity strain as an internal parameter. Many simplified models of stress-strain relationship have been established and applied in industry. But, for accurate metal forming, the very small elastic recovery stress may produce significant error respect with the desired geometric configuration. This problem is caused by the theoretic form of finite deformation.

To solve the theoretic problem of finite deformation, Trusedell's polar decomposition of deformation gradient[1] is widely used in large deformation problem. However, the orthogonal rotation in Trusedell's polar decomposition does not represent the real relative local rotation of material point respect with its neighbouring material[8], as by experimental observation the relative local rotation and the local stretching are happened in the same time rather than in consequence. Hence the rotation defined by Trusedell's polar decomposition only represents the equivalent orientation of deformed material element. This problem can be solved by Eringen's polar decomposition through discarding the symmetry requirement of stretching tensor[4-5], this leads to the introduction of micropolar Lagrangean (or Eulerian) strain and curvature tensor[6]. Recently, based on the Eringen's polar decomposition, the micropolar rotation tensor is introduced into plasticity theory[6]. So, the orthogonal rotation tensor do has intrinsic mechanical significance.

Although the above advancement is significant in theoretic consideration, they do not give out an explicit formulation of rotation tensor by displacement field. This paper will shows that this shortage, in essential sense, can be overcome by Chen's additive decomposition of deformation gradient[7-10].

Based on the additive decomposition of deformation gradient, a theory of finite deformation is established by Chen Z D. In his theory, the average local rotation gets clear geometric and mechanics meaning and an intrinsic reference is taken to express the relation between the current configuration and the initial configuration[9-10]. Unfortunately, the value of Chen's additive decomposition of deformation gradient for plasticity has not been well developed[10]. The main problem is that for large shear Chen's definition of the average local rotation requires a too strict condition. Fortunately, this problem has been solved recently by further research in that Chen's additive decomposition is extended to arbitrary shear[11]. The related results will be simply introduced in this paper.

Generally, a complete deformation process will produce irreversible deformation (such as plastic deformation). When the loading is completely removed, the residual deformation forms a deformed geometric configuration. For the unloading process, the recoverable deformation is elastic. The path-dependent feature of constitutive relation between stress and strain can be naturally formulated by Chen's strain definition. For the maximum loading process, the deformation process will evolutes from elastic, through plastic, to flow. This process is strictly formulated with the help of motion equations and deformation geometry. When the dynamic loading is removed, the rheology feature is explicitly formulated by taking the original geometry of material as the reference configuration. Hence, for a complete deformation from elastic behavior to flow, this paper explicitly formulates the related fields by displacement gradient field and invariant elastic constants.

## 2 Chen's SR Additive Decomposition of Deformation Gradient and Its Extension

For continuum, the displacement gradient can be divided into symmetry part and asymmetry part in classical treatment. For infinitesimal deformation, the asymmetry part is interpreted as rotation after Stokes. However, for large deformation this asymmetry part can not represent the true local rotation. Chen Z D shows that the displacement gradient can be divided into symmetry part and orthogonal rotation part. Hence, the true local rotation can be defined and be given its true geometric interpretation.

Based on the concept of points set transformation, Chen Z D introduces the concept of base vectors transformation between initial configuration and present configuration of continuum [7-10]. Taking co-moving dragging coordinator system, the transformation tensor $F_j^i$ is defined as:

$$\vec{g}_j = F_j^i \vec{g}_i^0 \tag{1}$$

Where, $\vec{g}_i$ and $\vec{g}_i^0$ are base vectors for present configuration and original configuration, respectively. The points set is coordinated with Lagrangean coordinators, here anti-covariant components $x^i, i=1,2,3$ are used. Note that the co-moving requires both base vectors for present configuration and original configuration be defined on the continuum. The rigid rotation or motion will not change such a relation. The dragging requires both base vectors be parameterized by the same anti-covariant components. So, such a definition of coordinator system is an intrinsic definition. This point is very important to understand the concept of Chen's point set transformation and its geometric meaning.

The transformation tensor can be expressed by the gradient of displacement $u^i$ as following:

$$F^i_j = u^i\big|_j + \delta^i_j \tag{2}$$

Where, $u^i\big|_j$ express the covariant derivation of displacement fields; $\delta^i_j$ is Kronecker-delta.

Chen Z D has shown that the transformation can be decomposed into the addition of one symmetry tensor expressing stretching and one unit orthogonal tensor expressing local rotation [7-10]. That is, we have:

$$F^i_j = S^i_j + R^i_j \tag{3}$$

Where:

$$S^i_j = \frac{1}{2}(u^i\big|_j + u^i\big|_j^T) - (1-\cos\Theta)L^i_k L^k_j \tag{4}$$

$$R^i_j = \delta^i_j + \sin\Theta \cdot L^i_j + (1-\cos\Theta)L^i_k L^k_j \tag{5}$$

$$L^i_j = \frac{1}{2\sin\Theta}(u^i\big|_j - u^i\big|_j^T) \tag{6}$$

$$\sin\Theta = \frac{1}{2}[(u^1\big|_2 - u^2\big|_1)^2 + (u^2\big|_3 - u^3\big|_2)^2 + (u^3\big|_1 - u^1\big|_3)^2]^{\frac{1}{2}} \tag{7}$$

In above expressions, the upper T represents transpose of a tensor, the parameter $\Theta$ is called local average rotation angel (from continuity consideration we can set $|\Theta| < \pi/2$) and tensor $L^k_j$ defines the local average rotation axis direction, it is an anti-symmetric tensor. The $S^i_j$ is a symmetric tensor (it is defined as strain by Chen Z D), the $R^i_j$ is a proper orthogonal tensor. They are two-point tensor [1,15], which has been discussed by Trusedell in details. To make the two-point tensor clear, the upper index corresponds to the base in initial configuration and the lower index corresponds to the base in current configuration. Mathematically, Chen Z D has proved that the upper index corresponds to anti-covariant components and the lower index corresponds to covariant components with respect to the base vectors in initial reference configuration. (A more strictly mathematical discussion is given in Appendix A).

But, the definition of local rotation angel is too strong to be reasonable for large shear deformation, as it requires the condition of:

$$\frac{1}{2}[(u^1\big|_2 - u^2\big|_1)^2 + (u^2\big|_3 - u^3\big|_2)^2 + (u^3\big|_1 - u^1\big|_3)^2]^{\frac{1}{2}} \leq 1 \tag{8}$$

This problem can be overcome by redefining the local average rotation angel $\theta$ (from continuity consideration we can set $|\theta| < \pi/2$) as the following [11]:

$$(\cos\theta)^{-2} = 1 + \frac{1}{4}[(u^1\big|_2 - u^2\big|_1)^2 + (u^2\big|_3 - u^3\big|_2)^2 + (u^3\big|_1 - u^1\big|_3)^2] \tag{9}$$

Redefining the local average rotation axis direction tensor $\tilde{L}^i_j$ as:

$$\tilde{L}^i_j = \frac{\cos\theta}{2\sin\theta}(u^i\big|_j - u^i\big|_j^T) \tag{10}$$

It is easy to verify that Chen's S-R additive decomposition theorem can be extended as [11]:

$$F^i_j = \tilde{S}^i_j + (\cos\theta)^{-1}\tilde{R}^i_j \tag{11}$$

Where:

$$\tilde{S}^i_j = \frac{1}{2}(u^i\big|_j + u^i\big|_j^T) - (\frac{1}{\cos\theta} - 1)(\tilde{L}^i_k \tilde{L}^k_j + \delta^i_j) \tag{12}$$

$$(\cos\theta)^{-1}\tilde{R}^i_j = \delta^i_j + \frac{1}{2}(u^i\big|_j - u^i\big|_j^T) + (\frac{1}{\cos\theta} - 1)(\tilde{L}^i_k \tilde{L}^k_j + \delta^i_j) \tag{13}$$

$$\tilde{R}^i_j = \delta^i_j + \sin\theta \cdot \tilde{L}^i_j + (1-\cos\theta)\tilde{L}^i_k \tilde{L}^k_j \tag{14}$$

Note that when the condition (8) exists (that is to say when $|\theta| \leq \pi/4$) there are two forms of Chen's S-R additive decomposition. They represent two possible deformations. There corresponding mechanical implication will discussed in next section when it is appropriate. Hereafter, the paper will name them as Chen's form-one (defined by equation (3)) and Chen's form-two (defined by equation (11)), respectively. This feature makes Chen's additive decomposition be explicitly used to express pure static deformation and elastic-plastic deformation, even to flow motion.

## 3 Strain Relation Between Additive Decomposition and Polar Decomposition

Based on Eringen's polar decomposition theorem[4-6], we have:

$$F_j^i = \tilde{R}_j^k \tilde{U}^{ik} = \tilde{V}_{kj} \tilde{R}_k^i \tag{15}$$

Comparing with Chen's additive decomposition form-two (11):

$$F_j^i = \tilde{S}_j^i + (\cos\theta)^{-1} \tilde{R}_j^i = \tilde{R}_j^k [\tilde{S}_l^i \tilde{R}_l^k + (\cos\theta)^{-1} \delta^{ik}] = [\tilde{S}_j^l \tilde{R}_k^l + (\cos\theta)^{-1} \delta_{jk}] \tilde{R}_k^i \tag{16}$$

We find that the pseudo-right stretch and pseudo-left stretch tensors are:

$$\tilde{U}^{ij} = \tilde{S}_k^i \tilde{R}_k^j + (\cos\theta)^{-1} \delta^{ij} \tag{17}$$

$$\tilde{V}_{ij} = \tilde{S}_j^k \tilde{R}_i^k + (\cos\theta)^{-1} \delta_{ij} \tag{18}$$

In form, we can introduce micropolar Lagrangean right-strain $^{Right}\tilde{\varepsilon}^{ij}$ tensor and left-strain tensor $^{Left}\tilde{\varepsilon}_{ij}$, respectively defined by pseudo-right stretch and pseudo-left stretch strain tensors[6] $^{Right}\tilde{\varepsilon}^{ij} = \tilde{U}^{ij} - \delta^{ij}$ and $^{Left}\tilde{\varepsilon}_{ij} = \tilde{V}_{ij} - \delta_{ij}$.

Based on the symmetry of $\tilde{S}_j^i$ tensor, viewing that the pseudo-left stretch tensor corresponds to the base vector orientation of initial configuration firstly rotated to current configuration then followed pure deformation, and viewing that the pseudo-right stretch tensor corresponds to suffered firstly a pure deformation then the base vector orientation of current configuration rotated to initial configuration, we get the definition of name strains:

$$\tilde{\varepsilon}^{ij} = \tilde{S}_k^j \tilde{R}_k^i + (\frac{1}{\cos\theta} - 1)\delta^{ij} \tag{19}$$

$$\tilde{\varepsilon}_{ij} = \tilde{S}_i^k \tilde{R}_j^k + (\frac{1}{\cos\theta} - 1)\delta_{ij} \tag{20}$$

The $\tilde{\varepsilon}^{ij}$ is the name strain defined in the base vector orientation of the initial reference configuration, the $\tilde{\varepsilon}_{ij}$ is the name strain defined in the base vector orientation of the current reference configuration. All components of tensors are explicitly expressed by the displacement gradient field. Note that they are not symmetric tensors, generally.

Note that for Chen's form-one, following similar discussion, we have:

$$\varepsilon^{ij} = S_k^j R_k^i \tag{21}$$

$$\varepsilon_{ij} = S_i^k R_j^k \tag{22}$$

The advantage of defining the name strains in such a way is that the name stress field and the name strain field are used widely in experiments and most constitutive parameters are calculated by them. From theoretical consideration, such a definition of name strain overcomes the ambiguity of strain definition in polar decomposition (such as micropolar Lagrangean strain[5-6], Cauchy-Green strain[15]) caused by the feature of two-point tensor. In fact, the ambiguity of strain definition in polar decomposition damaged the theoretic development and application as they cause many misunderstanding and controversy, even strong debate[9-10].

Now we turn to discuss how to express right Cauchy-Green deformation tensor and left Cauchy-Green deformation tensor, which are based on Trusedell's polar decomposition theorem[15].

Firstly, we note that the gauge tensor for current configuration is:

$$g_{ij} = \vec{g}_i \cdot \vec{g}_j = F_i^l \vec{g}_l^0 \cdot F_j^k \vec{g}_k^0 = F_i^l F_j^k g_{lk}^0 \tag{21}$$

Hence, when standard Cartesian coordinate system is taken as the initial configuration (that is to set $g_{ij}^0 = \delta_{ij}$), the right Cauchy-Green deformation tensor should be defined as[1]:

$$C_{ij} = F_i^l F_j^l \tag{22}$$

Note that in essential sense, this leads the definition of strain in current configuration in different form, as the initial gauge field is taken as reference.

Then, put Chen's additive decomposition form-two (11) into it, we get:

$$C_{ij} = \tilde{S}_i^l \tilde{S}_j^l + \frac{1}{\cos\theta}(\tilde{S}_i^l \tilde{R}_j^l + \tilde{S}_j^l \tilde{R}_i^l) + \frac{1}{\cos^2\theta}\delta_{ij} \tag{23}$$

This equation gives out the explicit expression of right Cauchy-Green deformation tensor under Chen's additive composition. It shows that the local average rotation has significant contribution to Cauchy-Green deformation tensor. Therefore, in the multiplicative decomposition[15] ($C=F^TF=U^2$, $F=RU$) the proper orthogonal tensor does not represent the intrinsic rotation of material element, as the symmetry of U is a too strong mathematic-style requirement.

We also find that, when standard Cartesian coordinate system is taken as the initial configuration, the positive and symmetric gauge tensor of current configuration is:

$$g_{ij} = \tilde{S}_i^l \tilde{S}_j^l + \frac{1}{\cos\theta}(\tilde{S}_i^l \tilde{R}_j^l + \tilde{S}_j^l \tilde{R}_i^l) + \frac{1}{\cos^2\theta}\delta_{ij} \tag{24}$$

Note that it is defined on the current reference configuration.

Although based on above discussion, for taking standard Cartesian coordinate system as the co-moving coordinator system in initial configuration one will make the explicit expression (23) of right Cauchy-Green deformation tensor available by displacement gradient field, the explicit expression of right and left stretch tensors (introduced by Trusedell[15]) is not available, as for a positive symmetric tensor the proper orthogonal tensor is not unique. This problem has caused a long-lasting problem in the strain calculation and different understanding about the real physical meaning of rotation. Guo Zhonghen points out that as the Cauchy-Green deformation strain tensor is positive and symmetric, there exists principle axes to express the Cauchy-Green deformation strain tensor[12], later he established a "non-compatible theory of deformation".

Viewing the local average rotation corresponds to an additive rigid rotation, for physical admissible elastic deformation, Chen Zhida suggests using the $S_j^i$ in Chen's decomposition form-one (4) to define the elastic strain[9-10]. Based on resent research, to correctly formulate elastic wave phenomena, the strain would be defined by $F_j^i - \delta_j^i$ directly[11]. As Chen Z D has augured that, by experimental observation, the relative local average rotation and the local stretching are happened in the same time rather than in consequent time expressed by polar decomposition. The local average rotation has its significance which is completely different with the pure rigid rotation, although it is formed in such a way that it is an orthogonal tensor. To make this point clear, let's consider an infinitesimal deformation:

$$F_j^i = R_j^i \approx \frac{1}{2}(u^i|_j - u^i|_j^T) + \delta_j^i \tag{24-2}$$

In elastic dynamics, it describes the strain of pure-shear wave. So, the orthogonal tensors in Chen's additive decomposition have intrinsic mechanics meaning.

Based on research on the physical meaning of non-symmetric field[11,13-15], to correctly formulate finite deformation phenomena, the strain would be defined by $F_j^i - \delta_j^i$.

(1) For pure elastic deformation, formulated by Chen's decomposition form-one, strains are defined as:

$$\varepsilon_j^i = S_j^i = \frac{1}{2}(u^i|_j + u^i|_j^T) - (1-\cos\Theta)L_k^i L_j^k \tag{25}$$

$$w_j^i = R_j^i - \delta_j^i = \sin\Theta \cdot L_j^i + (1-\cos\Theta)L_k^i L_j^k \tag{26}$$

where, the first one is named as stretch-contraction strain, the second one is named as rotation strain as it compares the local variation of base vector produced by deformation. Their addition forms the total strain.

(2) For plastic deformation, formulated by Chen's decomposition form-two, name strains are defined as:

$$\tilde{\varepsilon}_j^i = \tilde{S}_j^i = \frac{1}{2}(u^i|_j + u^i|_j^T) - (\frac{1}{\cos\theta}-1)(\tilde{L}_k^i \tilde{L}_j^k + \delta_j^i) \tag{27}$$

$$\tilde{w}_j^i = (\cos\theta)^{-1}\tilde{R}_j^i - \delta_j^i = \frac{1}{2}(u^i|_j - u^i|_j^T) + (\frac{1}{\cos\theta}-1)(\tilde{L}_k^i \tilde{L}_j^k + \delta_j^i) \tag{28}$$

where, the first one is named as stretch-contraction strain, the second one is named as rotation-expansion strain as it not only compares the local variation of base vector produced by deformation but also the expansion accompanied during deformation. Their addition forms the total strain.

It is clear that the above four tensors are two-point tensor. The most significant advantage of defining tensor in this way is that they are additive and in linear form (although in essential sense they are not linear about displacement gradient field).

The name strain definition (19) and (20) can be expressed by above definition as (for Chen's

form-two):

$$\tilde{\varepsilon}^{ij} = \tilde{S}_k^j \tilde{R}_k^i + (\frac{1}{\cos\theta} - 1)\delta^{ij} = \frac{1}{\cos\theta}\tilde{\varepsilon}_k^i(\tilde{w}_k^j + \delta_k^j) + (\frac{1}{\cos\theta} - 1)\delta^{ij} \tag{29}$$

$$\tilde{\varepsilon}_{ij} = \tilde{S}_i^k \tilde{R}_j^k + (\frac{1}{\cos\theta} - 1)\delta_{ij} = \frac{1}{\cos\theta}\tilde{\varepsilon}_i^k(\tilde{w}_j^k + \delta_j^k) + (\frac{1}{\cos\theta} - 1)\delta_{ij} \tag{30}$$

It can be seen that the local average rotation parameter $\theta$ plays an important role in the name strain definition ((19) and (20)) based on Eringen's polar decomposition theorem.

For pure elasticity (Chen's form-one), following similar procedure as for $\tilde{\varepsilon}^{ij}$ and $\tilde{\varepsilon}_{ij}$, introducing name strain $\varepsilon^{ij}$ and $\varepsilon_{ij}$, we have:

$$\varepsilon^{ij} = S_k^j R_k^i = \varepsilon_k^i(w_k^j + \delta_k^j) \tag{31}$$

$$\varepsilon_{ij} = S_i^k R_j^k = \varepsilon_i^k(w_j^k + \delta_j^k) \tag{32}$$

Comparing equations (31) and (32) with equations (25) and (26), and comparing equation (29) and (30) with equations (27) and (28), we find that the name strains based on Eringen's polar decomposition theorem can be further decomposed as the multiplicative of stretch-contraction name strain and rotation(-expansion) name strain adding stretch-contraction name strain. This explains the significant disadvantage of the name strains based on Eringen's polar decomposition.

Therefore, for the plastic deformation, the name strain (27) and (28) will be used hereafter in this paper. For pure elastic deformation, the strain (25) and (26) will be used hereafter.

**4 How to Express Path-dependent Deformation in Plasticity**

In traditional linear elasticity theory, using the name strain definition (25), the deformation energy is defined as a scalar:

$$\Phi = \Phi(S_j^i) = \Phi(\varepsilon_j^i)$$

Hence, in pure elasticity, for isotropic material, the elastic name stress tensor can be defined as:

$$\sigma_j^i = \lambda \varepsilon_l^l \delta_j^i + 2\mu \varepsilon_l^i \delta_j^l = \lambda \varepsilon_l^l \delta_j^i + \mu(\varepsilon_l^i \delta_j^l + \varepsilon_j^l \delta_l^i) \tag{33}$$

For wave motion, it corresponds to P-wave motion.

The corresponding isotropic elasticity tensor in mix form[16] is expressed as:

$$C_{jl}^{ik} = \lambda \varepsilon_j^i \delta_l^k + 2\mu \varepsilon_l^i \delta_j^k \tag{34}$$

However, the deformation which meets $F_j^i = R_j^i \neq \delta_j^i$ do exist (by definition (3-7) there exist non-null deformation gradient). So the $R_j^i$ should be included in the definition of deformation energy[6,9-10]. In fact, the above deformation energy definition will require that $R_j^i = \delta_j^i$, so it requires that the deformation gradient is a symmetric tensor or discarding the local average rotation by introducing symmetric strain tensor[15]. Therefore, the deformation energy (also depends on the local average rotation of deformation) should be defined as the scalar:

$$\Phi = \Phi(F_j^i - \delta_j^i) = \Phi(S_j^i + R_j^i - \delta_j^i) = \Phi(\varepsilon_j^i + w_j^i)$$

So, the rotation name stress $\tau_j^i$ will be expressed as:

$$\tau_j^i = \lambda w_l^l \delta_j^i + 2\mu w_l^i \delta_j^l \tag{35}$$

For wave motion, it corresponds to S-wave motion.

It is easy to verify that the total name stress

$$\bar{\sigma}_j^i = \sigma_j^i + \tau_j^i = \lambda(u^l|_l)\delta_j^i + 2\mu(u^i|_l)\delta_j^l \tag{36}$$

just is the definition of deformation stress (Cauchy stress) widely used in fluid dynamics, such as in Navier-Stokes equations, and seismics. This justifies our stress definition with phenomenological and practical soundness.

As the rotation name strain is not a symmetric tensor, for initial isotropic material, the stress field will be not symmetric. The non-symmetry of name stress field has caused strong and long lasting interests. Here non-symmetric stress field will be used without further discussion. For clarity to see, we will still follow the rule that the upper index represents the initial reference configuration and the lower index represents the current reference configuration.

Geometrically, for the mix form stress, the lower index defines the surface normal direction, the upper index defines the surface component direction[9-10]. Viewing that the $g^{ij}$ defines the area gauge of the coordinator system in current configuration and that the $g_{ij}$ defines the length gauge of the coordinator system in current configuration, the mix form stress can be transformed into true stress.

The mix form stress can be transformed into the true stress taking base vector in orientation of initial configuration as:

$$\sigma^{ij} = \sigma_l^i g^{jl}, \quad \tau^{ij} = \tau_l^i g^{jl} \tag{37}$$

Or, be transformed into the true stress taking base vector in orientation of current configuration as:

$$\sigma_{ij} = \sigma_j^l g_{il}, \quad \tau_{ij} = \tau_j^l g_{il} \tag{38}$$

This topic has been discussed by Chen Z D in detail[9-10]. The main point to understand that the stress is defined on surface force, one index related with the actual area of the surface and another index related with component direction of the surface force.

Now we turn to consider the first loading process. Its deformation path will be named as initial path, hereafter.

For initial isotropic material, initially the first loading process will be pure elastic. As grain rotations do exist in metal plasticity[17], we can introduce parameter $\theta_c$ to define the maximum allowed elastic local average rotation. When $\theta \geq \theta_c$, the local average rotation will be irreversible. To take the grain rotation into consideration, we will use the Chen's decomposition form-two.

On the other hand, to describe the initial path, we can take the plastic deformation which is determined by the maximum loading as an intermediate reference configuration, which is defined as:

$$G_{ij} = F_{Pi}^l F_{Pj}^k g_{lk}^0, \quad \vec{G}_j = F_{Pj}^i \vec{g}_i^0 \tag{39}$$

The deformation (11), referring to the intermediate plastic reference configuration, will be described as:

$$\vec{g}_j = F_j^i \vec{g}_i^0 = (F_{Ej}^i + F_{Pj}^i)\vec{g}_i^0 = F_{Ej}^i \vec{g}_i^0 + \vec{G}_i \tag{40}$$

where:

$$F_{Ej}^i = S_{Ej}^i + (R_{Ej}^i - \delta_j^i) \tag{41}$$

It defines the pure elastic deformation, which is superposed on the plastic deformation. By this definition, the deformation is decomposed as the stack of elastic deformation and plastic deformation.

The definition equation (40) can be rewritten as:

$$\vec{g}_j = F_j^i \vec{g}_i^0 = F_{Ej}^i \vec{g}_i^0 + \vec{G}_i = [F_{Ej}^l (F_{Pl}^i)^{-1} + \delta_j^i]\vec{G}_i = \tilde{F}_{ej}^i \vec{G}_i \tag{42}$$

Hence, the additive decomposition is readily related with the multiplicative decomposition of elastic deformation and plastic deformation, which is widely used in plasticity theory.

The advantage of additive decomposition is that the displacement gradient field referring to initial configuration can be decomposed as the stack of elastic gradient and plastic gradient. While, the multiplicative decomposition views the elastic deformation ($\tilde{F}_{ej}^i$) be defined on the displacement gradient measured on plastic configuration ($G_{ij}$). In fact, the intrinsic difference between the additive decomposition and the multiplicative decomposition is that the former defines elastic deformation along the unloading (elastic) path, which takes the plastic configuration as the point of null elastic deformation, and the later defines elastic along the initial path, which takes the plastic configuration as present referent configuration.

As for the actual deformation different maximum loading produces different plastic configuration, the multiplicative decomposition is more convenient for treating the incremental deformation[16,17] along initial path. Physically, the multiplicative composition describes the first-loading deformation path while the additive decomposition describes the un-loading deformation path.

For unloading process, the elastic stress can be given by constitutive equation along unloading path as:

$$\sigma_{Ej}^i = \lambda F_{El}^l \delta_j^i + 2\mu F_{El}^i \delta_j^l \tag{43}$$

On the other hand, for the multiplicative decomposition, at each point on the initial path there is a corresponding plastic configuration, hence the elastic stress field can be completely described by taking the plastic deformation as internal parameter. This forms the classical treatment of plasticity. The key point of this treatment is that the deformation is physically continuous while the stress is path-dependent.

By equation (40) we have:

$$F_{Ej}^i = (\tilde{F}_{ej}^l - \delta_j^l) F_{Pl}^i \tag{44}$$

Hence, we have:
$$\sigma_{Ej}^{i} = \lambda(\tilde{F}_{ek}^{l} - \delta_{k}^{l})F_{Pl}^{k}\delta_{j}^{i} + 2\mu(\tilde{F}_{ek}^{l} - \delta_{k}^{l})F_{Pl}^{i}\delta_{j}^{k} \qquad (45)$$

On the other hand, by equation (42), for multiplicative decomposition, along the initial path, we have:
$$\tilde{\sigma}_{ej}^{i} = \tilde{\lambda}[\tilde{F}_{el}^{l} - 3]\delta_{j}^{i} + 2\tilde{\mu}(\tilde{F}_{el}^{i} - \delta_{l}^{i})\delta_{j}^{l} = \tilde{\lambda}F_{Ek}^{l}(F_{Pl}^{k})^{-1}\delta_{j}^{i} + 2\tilde{\mu}F_{El}^{k}(F_{Pk}^{i})^{-1}\delta_{j}^{l} \qquad (46)$$

Note that the $\tilde{\lambda}$, $\tilde{\mu}$ are incremental elasticity which depends on the plasticity point and, usually, the incremental elasticity is isotropic[18].

As the upper index represents the plastic reference configuration, it can be converted into the initial reference configuration:
$$\sigma_{ej}^{i} = \tilde{\sigma}_{ej}^{l}G_{kl}\delta^{ki} = [\tilde{\lambda}F_{Ek}^{l}(F_{Pl}^{k})^{-1}\delta_{j}^{m} + 2\tilde{\mu}F_{El}^{k}(F_{Pk}^{m})^{-1}\delta_{j}^{l}]G_{mn}\delta^{in} \qquad (47)$$

It is clear that this stress is much different with the stress defined by (43). This is the main feature of path-dependent deformation. In most case, plastic hardening (or softening) can be defined as:
$$\tilde{C}_{jl}^{ik} = [\tilde{\lambda}(F_{Pl}^{k})^{-1}\delta_{j}^{m} + 2\tilde{\mu}(F_{Pl}^{m})^{-1}\delta_{j}^{k}]G_{mn}\delta^{in} \qquad (48)$$

This is used to evaluate the maximum loading ability of the plastic material.

Introducing the maximum loading point, the two paths meet at this point (yielding point). Hence, we have stress condition: $\sigma_{Ej}^{i} = \sigma_{ej}^{i}$, that is:
$$\lambda F_{El}^{l}\delta_{j}^{i} + 2\mu F_{El}^{i}\delta_{j}^{l} = \tilde{\lambda}F_{Ek}^{l}(F_{Pl}^{k})^{-1}G_{mj}\delta^{im} + 2\tilde{\mu}F_{Ej}^{k}G_{mk}\delta^{im} \qquad (49)$$

and geometric condition:
$$F_{j}^{i} = F_{Ej}^{i} + F_{Pj}^{i} = \tilde{F}_{ej}^{l}F_{Pl}^{i} \qquad (50)$$

Studying equation (49), one will find:
$$C_{jl}^{ik} = \lambda\delta_{j}^{i}\delta_{l}^{k} + 2\mu\delta_{l}^{i}\delta_{j}^{k} = [\tilde{\lambda}(F_{Pl}^{k})^{-1}\delta_{j}^{m} + 2\tilde{\mu}(F_{Pl}^{m})^{-1}\delta_{j}^{k}]G_{mn}\delta^{in} = \tilde{C}_{jl}^{ik} \qquad (51)$$

For isotropic plasticity, let: $F_{Pj}^{i} = \frac{1}{\cos\theta_{P}}\delta_{j}^{i}$, we have: $\tilde{\lambda} = \cos\theta_{P}\cdot\lambda$, $\tilde{\mu} = \cos\theta_{P}\cdot\mu$. It shows that for multiplicative elasticity and plasticity, the incremental elasticity not only tends to be small but also dependents on the plasticity. This is the most significant disadvantage of such a kind of plastic theory. To overcome this theoretic problem, we need to introduce the intrinsic definition of Chen's strain definition as the next section will shows.

## 5 Formulating Plasticity in Intrinsic Frame

For plastic deformation, formulated by Chen's decomposition form-two, at the maximum loading point, the name strains are defined as:
$$\tilde{\varepsilon}_{j}^{i} = \tilde{S}_{j}^{i} = \frac{1}{2}(u^{i}\big|_{j} + u^{i}\big|_{j}^{T}) - (\frac{1}{\cos\theta} - 1)(\tilde{L}_{k}^{i}\tilde{L}_{j}^{k} + \delta_{j}^{i}) \qquad (27)$$

If we rewrite it as:
$$\tilde{\varepsilon}_{j}^{i} = s_{j}^{i} + \tilde{S}_{Pj}^{i} = \frac{1}{2}(u^{i}\big|_{j} + u^{i}\big|_{j}^{T})_{e} + \frac{1}{2}(u^{i}\big|_{j} + u^{i}\big|_{j}^{T})_{P} - (\frac{1}{\cos\theta} - 1)(\tilde{L}_{k}^{i}\tilde{L}_{j}^{k} + \delta_{j}^{i}) \qquad (52)$$

Then an artificial deformation can be introduced to describe plasticity. To this purpose, let us consider the case when the loading is removed. As when the loading is removed, the stretching stress is zero, we have:
$$\tilde{S}_{Pj}^{i} = \frac{1}{2}(u^{i}\big|_{j} + u^{i}\big|_{j}^{T})_{P} - (\frac{1}{\cos\theta} - 1)(\tilde{L}_{k}^{i}\tilde{L}_{j}^{k} + \delta_{j}^{i}) = 0 \qquad (53)$$

That is to say, after the maximum loading, the irreversible displacement gradient is completely attributed to the irreversible local average rotation.

Hence, for unloading from the maximum loading point, the name stress variant will be:
$$\Delta\sigma_{j}^{i} = \lambda s_{l}^{l}\delta_{j}^{i} + 2\mu s_{j}^{i} \qquad (54)$$

It forms a natural description for the unloading-path deformation.

On this sense, the plastic deformation is defined as:
$$F_{Pj}^{i} = \tilde{w}_{j}^{i} + \delta_{j}^{i} = \frac{1}{\cos\theta}\tilde{R}_{j}^{i} \qquad (55)$$

Based on above research, during the reloading process, once the deformation reaches the level of the maximum loading, the deformation will follow the initial path. The classical plastic name stress will be:

$$\sigma_{Pj}^i = \lambda \widetilde{w}_l^l \delta_j^i + 2\mu \widetilde{w}_j^i = 5\lambda(\frac{1}{\cos\theta} - 1)\delta_j^i + \frac{2\mu}{\cos\theta}\widetilde{R}_j^i \tag{56}$$

So, the classical plastic name stress can be determined by the three parameters which describe the plastic deformation.

Now, we consider the reloading path which follows the elastic path before it reaches the maximum loading point. Here, the incremental elasticity[18] will be determined by equation (51) in form:

$$\widetilde{\lambda} = \cos\theta_P \cdot \lambda, \quad \widetilde{\mu} = \cos\theta_P \cdot \mu \tag{57}$$

Hence, when the maximum elasticity range $\bar{s}_j^i$ is given by engineering consideration, the classical maximum loading stress $P_j^i$ can be approximated along the initial loading path as:

$$P_j^i = \cos\theta \cdot (\lambda \bar{s}_l^l \delta_j^i + \mu \bar{s}_j^i) + 5\lambda(\frac{1}{\cos\theta} - 1)\delta_j^i + \frac{2\mu}{\cos\theta}\widetilde{R}_j^i \tag{58}$$

On the other hand, when the maximum elasticity range $\bar{s}_j^i$ is given, the classical maximum loading stress $P_j^i$ can be calculated along the unloading (reloading) path directly as:

$$P_j^i = \lambda \bar{s}_l^l \delta_j^i + \mu \bar{s}_j^i \tag{59}$$

Finally, we get the relationship equation between the required plastic deformation and the un-required elastic recovery as:

$$(1 - \cos\theta) \cdot (\lambda \bar{s}_l^l \delta_j^i + \mu \bar{s}_j^i) = 5\lambda(\frac{1}{\cos\theta} - 1)\delta_j^i + \frac{2\mu}{\cos\theta}\widetilde{R}_j^i \tag{60}$$

Based on this equation, when the required plastic deformation is given, for the known material elasticity $(\lambda, \mu)$, the elastic deformation range $\bar{s}_j^i$ is determined. Consequently, the maximum loading stress is determined also by equations (58) or (59).

Therefore, by introducing an equivalent plastic deformation, the unloading path will get a natural description. In this way, the rotation stress will take the position of plastic stress (although it is introduced as the maximum loading stress).

Examining the equation (55), it will require that the plastic configuration gauge is isotropic. Mathematically, it is related with a conformal transformation defined by:

$$G_{ij} = F_{Pi}^l \vec{g}_l^0 \cdot F_{Pj}^k \vec{g}_k^0 = \frac{1}{\cos^2\theta} g_{ij}^0 \tag{59}$$

So, a broad class of plasticity can be described by this theoretic frame.

When the plasticity configuration is given, the equation (55) can be used to determine three independent parameters (needed for defining the rotation tensor) and, hence to determine the required maximum loading field by equations (58-60).

The main point in above formulation is to introduce the local average rotation as the irreversible deformation and based on the engineering strain definition to show that the Chen's strain definition contains the plasticity naturally, so the stress will naturally represent the path-dependent feature. For such an intrinsic formulation, only original elastic constants are needed to describe the whole process. In fact, for Chen's strain definition, the unloading process will lead to a special strain field ($\sigma_j^i = 0$) $\widetilde{S}_j^i = 0$, it will naturally introduce plasticity as the engineering (classical) strain is defined as: $\varepsilon_{ij} = \frac{1}{2}(u^i|_j + u^i|_j^T)$.

## 6 Motion Equation of Deformation

For the displacement field defined in the initial co-moving dragging coordinator system, the displacement or rotation of continuum as a rigid whole body has no contribution to the intrinsic deformation. So we only consider the conservation related with deformation. In the co-moving dragging coordinator system, the linear momentum conservation can be expressed in integral form as:

$$\frac{\partial}{\partial t}\int_\Omega (\rho \frac{\partial U^i}{\partial t})d\Omega = \oint_S \sigma_l^i n^l da \tag{60}$$

where, $n^l$ represent the unit normal vector of outside surface, $\rho$ is mass density at current configuration, $d\Omega$, $da$ represent the integration is taken at current configuration.

Based on Stokes Law, it is easy to find the differential form of linear momentum conservation:

$$(\sigma_l^i)\big|_l = \frac{\partial}{\partial t}(\rho \frac{\partial U^i}{\partial t}) \tag{61}$$

In the co-moving dragging coordinator system, the angular momentum conservation can be expressed in integral form as:

$$\frac{\partial}{\partial t}\int_\Omega (\vec{R}+\vec{U})\times(\rho\vec{V})d\Omega = \oint_S (\vec{R}+\vec{U})\times\vec{t}\,da \tag{62}$$

where, $V^i = \frac{\partial U^i}{\partial t}$, $\vec{R}$ is the initial position vector, $\vec{U}$ is the displacement vector, $t_j = \sigma_j^l n_l$, $n_l$ is the unit tangent vector of outside surface.

Note that $F_j^i = U^i\big|_j + \delta_j^i$, introducing extended Kronecker $e_{ijk}$, the equation can be expressed as:

$$\int_\Omega \sqrt{g^0}e_{ijk}(R^j+U^j)\frac{\partial}{\partial t}(\rho V^k)d\Omega = \int_\Omega \sqrt{g^0}e_{ijk}[(R^j+U^j)\sigma_k^m]\big|_m d\Omega \tag{63}$$

that is:

$$\int_\Omega \sqrt{g^0}e_{ijk}(R^j+U^j)\frac{\partial}{\partial t}(\rho V^k)d\Omega = \int_\Omega \sqrt{g^0}e_{ijk}F_m^j\sigma_k^m d\Omega + \int_\Omega \sqrt{g^0}e_{ijk}(R^j+U^j)(\sigma_k^m)\big|_m d\Omega \tag{64}$$

where, $g^0 = \det|g_{ij}^0|$。 Note that the velocity in initial configuration can be converted into velocity in current configuration as bellow:

$$V^i\vec{g}_i^0 = V^i g_{ij}^0 \vec{g}^{0j} = V^i g_{ij}^0 F_k^j \vec{g}^k \tag{65}$$

Hence, the differential form of angular momentum conservation is:

$$(\sigma_j^i)\big|_i = \frac{\partial}{\partial t}[(\rho\frac{\partial U^i}{\partial t})g_{il}^0 F_j^l] \tag{66}$$

$$e_{ijk}F_l^j\sigma_k^l = 0 \tag{67}$$

Therefore, the motion equation of deformation expressed by displacement field $U^i$ is:

$$(\sigma_l^i)\big|_l = \frac{\partial}{\partial t}(\rho\frac{\partial U^i}{\partial t}) \tag{68-1}$$

$$(\sigma_j^i)\big|_i = \frac{\partial}{\partial t}[(\rho\frac{\partial U^i}{\partial t})g_{il}^0 F_j^l] \tag{68-2}$$

$$e_{ijk}F_l^j\sigma_k^l = 0 \tag{68-3}$$

For large deformation, as the lower index of stress represents its component in current configuration, so the equation (69-2) is related with the velocity in current configuration.

If local average rotation is not considered, that is for the deformation:

$$F_j^i = S_j^i + \delta_j^i \tag{69}$$

The equation (68-3) will be met by a symmetric stress $\sigma_j^i$. If the stress $\sigma_j^i$ is symmetric, the equation (68-3) will require that the deformation $F_j^i$ must be symmetric. However, this can be true only for static deformation. Where, the static motion equations are:

$$(\sigma_l^i)\big|_l = 0 \tag{70-1}$$

$$(\sigma_j^i)\big|_i = 0 \tag{70-2}$$

$$e_{ijk}F_l^j\sigma_k^l = 0 \tag{70-3}$$

So, even for large deformation, the static stress must be symmetric, and hence, the static deformation must be symmetric. The conclusion is that the static mechanics of traditional deformation theory is correct. However, for dynamic deformation, it is correct only for infinitesimal deformation.

In classical infinitesimal deformation mechanics, $F_j^i \approx \delta_j^i$, so the motion equation (68) can be approximated as:

$$(\sigma_l^i)\big|_l = \frac{\partial}{\partial t}(\rho\frac{\partial U^i}{\partial t}) \tag{71-1}$$

$$(\sigma_j^i)\big|_i = \frac{\partial}{\partial t}(\rho \frac{\partial U^i}{\partial t} g_{ij}^0) \tag{71-2}$$

$$e_{ijk} F_l^j \sigma_k^l = 0 \tag{71-3}$$

In standard rectangular initial co-moving dragging coordinator system, it becomes:

$$(\sigma_l^i)\big|_l = \frac{\partial}{\partial t}(\rho \frac{\partial U^i}{\partial t}) \tag{72-1}$$

$$(\sigma_j^i)\big|_i = \frac{\partial}{\partial t}(\rho \frac{\partial U^i}{\partial t} \delta_{ij}) \tag{72-2}$$

$$e_{ijk} F_l^j \sigma_k^l = 0 \tag{72-3}$$

It is same as the traditional form, when the difference between initial configuration and current configuration is ignored. That is they can be written as:

$$\frac{\partial \sigma_{ij}}{\partial x^j} = \frac{\partial}{\partial t}(\rho \frac{\partial U^i}{\partial t}) \tag{73-1}$$

$$\sigma_{ij} = \sigma_{ji} = C_{ij}^{kl} \varepsilon_{kl} \tag{73-2}$$

$$\varepsilon_{ij} = \frac{1}{2}(\frac{\partial U^i}{\partial x^j} + \frac{\partial U^j}{\partial x^i}) \tag{73-3}$$

In fact, the equation (73) forms the bases of traditional elasticity theory.

To extend the traditional form (73) to large deformation, unfortunately, it was believed that the strain and stress should be rank-two symmetric co-variant tensor. Although Truessdell introduced two-point tensor to explain the feature of deformation gradient tensor, most of large deformation mechanics theory were based on extending the equation (73) under the condition of so-called covariant feature of strain and stress tensor[19-21]. Based on this research, the equation (73) is a too limited special case and, hence, can not be used as the rational bases for large deformation theory. This problem has attacked by different researchers from different view-points.

However, for large deformation, based on equation (68), the stress must mot be symmetric for dynamic deformation as the following equation must be met.

$$(\sigma_j^i - \sigma_j^{iT})\big|_i = \frac{\partial}{\partial t}[(\rho \frac{\partial U^i}{\partial t}) g_{il}^0 (F_j^l - \delta_j^l)] \tag{74}$$

Traditionally, this problem is addressed by non-linear elasticity theory where the elastic constants depend on stress or train, at the same time equation (73) are maintained. This research shows that such a way is only an approximation. The most significant shortage is the stress (or train) dependent elasticity parameters. This can not be made by a physical soundness interpretation.

For Chen's form-one, there are two typical deformations will be studded in the following sections. One is pure stretching deformation defined as: $F_j^i = S_j^i + \delta_j^i$, the motion equation is:

$$(\hat{\sigma}_l^i)\big|_l = \frac{\partial}{\partial t}(\rho \frac{\partial U_P^i}{\partial t}) \tag{75-1}$$

$$(\hat{\sigma}_j^i)\big|_i = \frac{\partial}{\partial t}[(\rho \frac{\partial U_P^i}{\partial t}) g_{il}^0 (S_j^l + \delta_j^l)] \tag{75-2}$$

$$e_{ijk}(S_l^j + \delta_l^j) \hat{\sigma}_k^l = 0 \tag{75-3}$$

It corresponds to a pure P-wave motion. $\hat{\sigma}_j^i = E_{jl}^{ik} S_k^l$. The existence of non-symmetric stress means that the strong P-wave motion will produce S-wave. As an approximation, if the S-wave displacement is ignored, to make the equation meaningful, anisotropic elasticity parameters can be introduced. As there are nine independent equations, the three displacement and six strain component are solvable for a given elasticity constants. That is to say, this equation is in closed-form.

Another one is pure orthogonal local rotation, defined as: $F_j^i = R_j^i, S_j^i = 0$, the motion equation is:

$$(\tilde{\sigma}_l^i)\big|_l = \frac{\partial}{\partial t}(\rho \frac{\partial U_S^i}{\partial t}) \tag{76-1}$$

$$(\tilde{\sigma}_j^i)\big|_i = \frac{\partial}{\partial t}[(\rho \frac{\partial U_S^i}{\partial t}) g_{il}^0 R_j^l] \tag{76-2}$$

$$e_{ijk} R_l^j \tilde{\sigma}_k^l = 0 \tag{76-3}$$

It corresponds to a pure S-wave motion. $\tilde{\sigma}_j^i = E_{jl}^{ik}(R_k^l - \delta_k^l)$. As $S_j^i = 0$ means that:

$$\frac{1}{2}(U^i\big|_j + U^i\big|_j^T) = (1-\cos\Theta) L_k^i L_j^k \tag{77}$$

So, the S-wave will produce P-wave motion. The most impotent feature is that two independent S-wave motion can exist as the equation (76-1) and (76-2) represent two different phase velocity of S-wave. This phenomenon has been observed and is named as Shear Wave Splitting by Stuart Crampin (Prof., Edinburgh University)[22-24]. For S-wave motion, the equation (76-3) means the elasticity constant tensor must be invariant under orthogonal rotation. It implies that for some special direction, if the elasticity constant tensor is invariant under orthogonal rotation, the S-wave cannot exist (or say propagate) in theses directions (named as S-wave inhibit direction in available document). This topic has been addressed by several researchers.

As the equation (76-3) set limits on the elasticity constants, it is trivial one. The six equations (76-1) and (76-2) can used to solve the three displacement and three independent parameters of orthogonal rotation. Hence, the equation is in close-form.

By Chen's form-two, there exists a new wave type, here named it as cracking wave. It is defined by the deformation: $F_j^i = \frac{1}{\cos\theta} R_j^i, S_j^i = 0$. For such a deformation, the motion equation is:

$$(\tilde{\sigma}_l^i)\big|_l = \frac{\partial}{\partial t}(\rho \frac{\partial U_S^i}{\partial t}) \tag{78-1}$$

$$(\tilde{\sigma}_j^i)\big|_i = \frac{\partial}{\partial t}[\frac{1}{\cos\theta}(\rho \frac{\partial U_S^i}{\partial t}) g_{il}^0 R_j^l] \tag{78-2}$$

$$e_{ijk} R_l^j \tilde{\sigma}_k^l = 0 \tag{78-3}$$

Where, $\tilde{\sigma}_j^i = E_{jl}^{ik}(\frac{1}{\cos\theta} R_k^l - \delta_k^l)$. Similar with the S-wave motion, the equation (78-3) sets limits on the elasticity constant tensor, so there exists inhibit direction for cracking wave. For a fixed $\theta$, the difference from S-wave is that the cracking wave speed is bigger than the S-wave speed. Further more, the bigger the $\theta$ is, the fast the wave speed is. Hence, the cracking wave speed lay in a wide range from S-wave speed to infinite (theoretically). The existence of inhibit direction for cracking wave means that by carefully setting the elasticity constants a reliable non-destructive direction can be made. This has important application in industry.

## 7 Fatigue Evolution of Pure Stretching Deformation

At high symmetry stress level, the deformation can be approximated by symmetric pure stretching deformation. That is:

$$F_j^i \approx S_j^i + \delta_j^i \tag{79}$$

Its motion equation (75) can be written as:

$$(\hat{\sigma}_j^i - \hat{\sigma}_j^{iT})\big|_i = \frac{\partial}{\partial t}(\rho \frac{\partial U^i}{\partial t} S_j^i) \tag{80}$$

$$(\hat{\sigma}_j^i + \hat{\sigma}_j^{iT})\big|_i = \frac{\partial}{\partial t}[(\rho \frac{\partial U^i}{\partial t})(S_j^i + 2\delta_j^i)] \tag{81}$$

For infinitesimal rotation deformation, the asymmetric stress is:

$$^{anti}\sigma_j^i = (\hat{\sigma}_j^i - \hat{\sigma}_j^{iT}) = \mu(U^i\big|_j - U^i\big|_j^T) \tag{82}$$

Suppose that the displacement field is harmonic:

$$U^i = U_0^i \cdot \cos(\omega t) \tag{83}$$

where, the lower index 0 means the amplitude of displacement motion, $\omega$ is the frequency of harmonic motion. Note that the stress and strain will have the same harmonic form. Hence, putting equation (83) into equation (80), it is found that:

$$(\hat{\sigma}_j^i - \hat{\sigma}_j^{iT})\big|_i = [E_{jl}^{ik}(\frac{\partial U^l}{\partial x^k} - \frac{\partial U^l}{\partial x^k}^T)]\big|_i = -\rho\omega^2 \cos(2\omega t) \cdot U_0^i S_{0j}^i \tag{84}$$

Combining with equation (82), it becomes:

$$\cos(\omega t) \cdot \left(\frac{\partial U_0^i}{\partial x^j} - \frac{\partial U_0^i}{\partial x^j}^T\right)\bigg|_i = -\frac{\rho \omega^2}{\mu} \cos(2\omega t) \cdot U_0^i S_{0j}^i \approx -\frac{\rho \omega^2}{\mu} \cos(2\omega t) \cdot \frac{\partial (U_0^i U_0^i)}{\partial x^j} \quad (85)$$

Note that for Chen's form-one, $L_j^i = \frac{1}{2\sin\Theta}(u^i|_j - u^i|_j^T)$, putting this definition into above equation, one will get:

$$\cos(\omega t) \cdot \frac{\partial}{\partial x^i}(\sin\Theta \cdot L_j^i) \approx -\frac{\rho \omega^2}{\mu} \cos(2\omega t) \cdot \frac{\partial (U_0^l U_0^l)}{\partial x^j} \quad (86)$$

For a given harmonic displacement deformation, this equation can be used to determine the infinitesimal rotation deformation. In mechanic dynamics, this equation can be used to explain the self-cite vibration or noise related with dynamic deformation body. Note that $L_j^i$ only has two independent components. So, only three independent parameters are to be determined, they are $\Theta$ and $L_j^i$.

Its typical solution is in form:

$$\cos(\omega t) \cdot (\sin\Theta \cdot L_j^i) \approx -\frac{\rho \omega^2}{\mu} \cos(2\omega t) \cdot (U_0^l U_0^l) + C \quad (87)$$

The existence of a time related global parameter $C$ means that, after a finite time of deformation, even the displacement field is zero, there is a time related residual local average rotation, which should be determined by initial and boundary conditions.

This gives the theoretic interpretation for the fatigue-cracking phenomenon in mechanical engineering. Note that the frequency in right side of equation (86) is two times of the frequency in left side of it. Further more, the left side and the right side are not in-phase. Hence it means that for a general dynamic process the equation (80) is a highly non-linear equation.

The equation (81) can be approximated as:

$$\frac{1}{2}(\hat{\sigma}_j^i + \hat{\sigma}_j^{iT})\bigg|_i = \frac{\partial}{\partial t}(\rho \frac{\partial U^i}{\partial t} \delta_j^i) \quad (88)$$

So, for average symmetry stress, the classical motion equation is a good approximation. When the maximum symmetry deformation is obtained by this equation, with suitable initial or boundary condition, the equation (86) can be used to obtain the residual local rotation. Hence, get a fatigue solution.

Putting equation (83) into it, one gets:

$$\frac{1}{2}(\hat{\sigma}_j^i + \hat{\sigma}_j^{iT})\bigg|_i = \frac{\partial}{\partial t}(\rho \frac{\partial U^i}{\partial t} \delta_j^i)$$

To make this equation be correct, the average strain should be modified by residual local average rotation:

$$S_j^i = \frac{1}{2}(U^i|_j + U^i|_j^T) - (1-\cos\Theta)L_k^i L_j^k \quad (89)$$

The average stress should be defined as:

$$\sigma_j^i = E_{jl}^{ik} S_k^l = \frac{1}{2} E_{jl}^{ik}(U^l|_k + U^l|_k^T) - E_{jl}^{ik}(1-\cos\Theta)L_m^l L_k^m \quad (90)$$

It can be concluded that the residual local average rotation will make the average stress decrease.

After a long dynamic process, when the continuum returns to zero stress state: $\sigma_j^i = 0$, by equation (90), one will get: $S_j^i = 0$, that means:

$$\frac{1}{2}(U^i|_j + U^i|_j^T) = (1-\cos\Theta)L_k^i L_j^k \quad (91)$$

hence, the fatigue-cracking effect can be defined by:

$$\varepsilon_{Pj}^i = (1-\cos\Theta)L_k^i L_j^k \quad (92)$$

Its macro-phenomenon can be explained as residual plastic deformation.

Geometrically, when $\Theta \to \pi/2$, such a deformation is physical impossible. That means it is the absolute destruction of the continuum. In fact, for a given material, there exists a critical $\Theta_{critical}$, which is determined by the intrinsic feature of material. When:

$$\Theta \geq \Theta_{critical} \quad (93)$$

the material will be destructive. So, the maximum strain load of the material can be estimated as:

$$^{Max}\varepsilon^i_j = \frac{1}{2}(U^i\big|_j + U^i\big|_j^T) \approx (1-\cos\Theta_{Critical})L^i_k L^k_j \quad (94)$$

Note that, for the evolution of pure stretching deformation, the residual local average rotation is not anisotropic in general case. When a fatigue-cracking effect happened along a direction, such a direction will become the profound direction for the further development of fatigue-cracking effect, until the material cracked plane appeared. The normal of material cracked plane just is the local rotation axe. After the cracking, the material will flow locally. This topic will be addressed in later sections.

In classical treatment of material fatigue-cracking effect, effective elastic constants are introduced to describe its effects on stress-strain relationship[25]. Here, the research gives a clear causal formulation.

## 8 Fatigue Evolutions for Pure Rotational Deformation

In low symmetry stress level, the deformation can be approximated by pure rotation deformation. That is:

$$F^i_j \approx R^i_j \quad (95)$$

For such a kind of deformation, the current gauge filed is same as the initial gauge field as $g_{ij} = R^l_i R^l_j = \delta_{ij}$ holds. In dynamic case, the motion equation is:

$$(\tilde{\sigma}^i_l)\big|_l = \frac{\partial}{\partial t}(\rho\frac{\partial U^i_S}{\partial t}) \quad (96\text{-}1)$$

$$(\tilde{\sigma}^i_j)\big|_i = \frac{\partial}{\partial t}[(\rho\frac{\partial U^i_S}{\partial t})R^i_j] \quad (96\text{-}2)$$

$$e_{ijk}R^j_l \tilde{\sigma}^l_k = 0 \quad (96\text{-}3)$$

The third equation requires that:

$$\tilde{\sigma}^i_j = 2\mu R^i_j \quad (97)$$

For an initial isotropic material, the constitutive equation is:

$$\tilde{\sigma}^i_j = \lambda(R^l_l - 3)\delta^i_j + 2\mu(R^i_j - \delta^i_j) \quad (98)$$

That is:

$$\lambda(R^l_l - 3)\delta^i_j = 2\mu\delta^i_j \quad (99)$$

As the equation (5) gives out:

$$R^l_l - 3 = (1-\cos\Theta)L^i_l L^l_i = 2(1-\cos\Theta) \quad (100)$$

Hence one gets:

$$\mu = (1-\cos\Theta)\cdot\lambda \quad (101)$$

For an infinitesimal rotation:

$$\Theta = \Theta_0 \cdot \cos(\omega t) \quad (102)$$

The effective dynamic shear constant is:

$$\mu = (1-\cos\Theta)\cdot\lambda \approx \frac{\lambda}{2}\Theta_0^2 \cdot \cos^2(\omega t) \quad (103)$$

So, for pure rotation deformation, the effective is rotation angular dependent. In this case, the traditional symmetry strain and stress are, respectively

$$\varepsilon^i_j = \frac{1}{2}(U^i\big|_j + U^i\big|_j^T) = (1-\cos\Theta)L^i_k L^k_j \quad (104)$$

$$\sigma^i_j = 2(1-\cos\Theta)\cdot[\lambda\delta^i_j + \mu L^i_k L^k_j] \quad (105)$$

Combining with equation (103), one gets:

$$\sigma^i_j \approx 2\lambda(1-\cos\Theta)\cdot[\delta^i_j + (1-\cos\Theta)L^i_k L^k_j] \approx 2\lambda(1-\cos\Theta)\cdot\delta^i_j \quad (106)$$

It shows that the dynamic process of pure rotation deformation will cause fatigue deformation also, but such a deformation cannot be observable in macro point as the gauge field has no variation.

The existence of such a kind of fatigue during pure rotational deformation can be approximated by the concept of residual stress. The larger the maximum rotation angular is, the larger the residual stress is.

Such a stress behaves as a positive pressure, which is an outward direction isotropic stress.

Note that when equation (97) holds, the motion equation be approximated as:

$$(\tilde{\sigma}_l^i)\big|_l = \frac{\partial}{\partial t}(\rho \frac{\partial U_S^i}{\partial t}) \quad (107\text{-}1)$$

$$(\tilde{\sigma}_j^i)\big|_i = \frac{\partial}{\partial t}[(\rho \frac{\partial U_S^i}{\partial t})R_j^i] \quad (107\text{-}2)$$

Where, the rotational stress (97) is used. Three parameters of rotation tensor and three displacements are to be determined. The equation is in close form. When the maximum local rotation solution is obtained, the equation (104) and (105) give out the fatigue solution.

From its very small shear constant, it can be conclude that such a material is shear-soft initially, and after long time duration of rotational deformation, the material will becomes harder to be sheared. This feature is significant when the rotation is large.

## 9 Cracking Evolution Condition

Generally, for infinitesimal deformation, the deformation is elastic. When the stress is large enough, plastic deformation will appear. When the dynamic process is long acting on the material, the fatigue will happen. So, there is a intrinsic parameter $\Theta_c$ to express the critical local rotation angular of material. This section will develop the related evolution equations for cracking.

When the local average rotation angular is bigger than the critical angular of material, the deformation gradient will take the Chen's form-two, that is:

$$F_j^i = \tilde{S}_j^i + (\cos\theta)^{-1}\tilde{R}_j^i \quad (11)$$

When cracking happens, the intrinsic stress will be zero, that make the stretching strain becomes zero. In this case, the deformation can be approximated as:

$$F_j^i = (\cos\theta)^{-1}\tilde{R}_j^i \quad (108)$$

Hence, the current gauge field will be:

$$g_{ij} = (\cos\theta)^{-2} g_{ij}^0 \quad (109)$$

Therefore, the cracking of material behaves as isotropic expansion. This expansion phenomenon during cracking is well reported by a lot of experiments.

At the critical cracking point, the deformation can be decomposed in two forms. They must be the same. Hence, the following condition must be met:

$$(\frac{1}{\cos\theta}-1)(\tilde{L}_k^i\tilde{L}_j^k + \delta_j^i) = (1-\cos\Theta_c)L_k^i L_j^k \quad (110)$$

Note that from equation (9), there is a relation between the two local rotation angular:

$$(\cos\theta)^{-2} = 1+\sin^2\Theta_c \quad (111)$$

The equations (110) and (111) determine the cracking condition in deformation gradient form.

When the Chen's form-one is known, for a given critical rotation angular, the cracking behavior is analytically solvable. For the inverse problem, when the cracking parameters are measured, the equation can be used to get the local rotation before the cracking happened.

There is a striking conclusion that the rotation axe of cracking is always different from the rotation axe of deformation before cracking.

Geometrically, no matter what material is, the absolute cracking will happens if $\Theta = \pi/2$. In such a cracking process,

$$(\cos\theta)^{-2} = 1+\sin^2\Theta_c = 2 \quad (112)$$

That gives a solution:

$$\theta_c = \pm\pi/4 \quad (113)$$

Such a kind of cracking forms a new gauge field: $g_{ij} = 2g_{ij}^0$, geometrically, it exhibits as two orthogonal planes its normal have an angular $\pi/4$ with the principle stress axe as the principle axes have an complete ($\Theta = \pi/2$) orthogonal rotation.

This sets a limit for the deformation of continuum:

$$-\pi/4 \le \theta_c \le \pi/4 \quad (114)$$

This condition corresponds to condition in Chen's form-one:

$$-\pi/2 \le \Theta \le \pi/2 \quad (115)$$

Only within this condition, the continuum deformation is elastic, plastic, or a more complicated form. If this condition does not be met again, the deformation becomes flow, which will soon be studied in next section.

Generally, for most kind of material the critical angular of material is less than $\pi/2$, so we conclude that for general case, the local rotation angular at the cracking point $\theta_c$ is less than $\pi/4$. In this case, the cracking planes will form angular $2\theta_c$, and $\pi - 2\theta_c$. For many rock materials, a $\theta_c = \pi/3$ cracking is observed. For this case, the $\Theta_c$ is $\arcsin(\sqrt{3}/3)$, it is bigger than $\pi/6$.

For these two typical cracking cases, the shear constant after cracking evolution are:
$$\mu = (1-\cos\Theta)\cdot\lambda = (1-\sqrt{2}/3)\lambda, \text{ for } \Theta_c = \pi/6$$
$$\mu = (1-\cos\Theta)\cdot\lambda = (1-\sqrt{2}/2)\lambda, \text{ for } \Theta_c = \pi/4$$
$$\mu = (1-\cos\Theta)\cdot\lambda = \lambda, \text{ for } \Theta_c = \pi/2$$

Therefore, after cracking, the shear constant in static state will be completely determined by the material critical angular and its initial stretching parameter $\lambda$.

Now, we turn to study the stress condition for cracking evolution. To this purpose, it should be point out that no matter what kind of deformation, it is the dynamic motion equations control the process before the cracking happened. As the dynamic process of deformation always require the asymmetry stress do exist, one can say even the deformation is symmetric, the requirement of dynamic motion will ask a local rotation. Hence, it should not be surprised that stress condition of cracking can be always be formulated with local rotation angular parameters.

At the cracking point, the classical strain is:
$$\varepsilon^i_j = \frac{1}{2}(U^i\big|_j + U^i\big|_j^T) = (1-\cos\Theta_c)L^i_k L^k_j \tag{116}$$

The classical stress is:
$$\sigma^i_j = E^{ik}_{jl}\varepsilon^l_k = E^{ik}_{jl}(1-\cos\Theta_c)L^l_m L^m_k \tag{117}$$

Define the yield stress as:
$$\sigma_S = E^{ik}_{jl}(1-\cos\Theta_c)\delta^l_m \delta^m_k \tag{118}$$

Then the yield condition becomes:
$$\sigma^i_j = \sigma_S L^l_m L^m_k \tag{119}$$

In principle stress space, it forms a sphere which is named as stress-sphere in classical yield theory. To make its relation with Tresca condition, Mises condition, and other forms of yield conditions[3,25] clear, it is appropriate to give a simple example.

Without lose of generality, consider the case of rotation axe is along $x^3$ coordinator:
$$L^i_j = \begin{vmatrix} 0 & L_3 & -L_2 \\ -L_3 & 0 & L_1 \\ L_2 & -L_1 & 0 \end{vmatrix} = \begin{vmatrix} 0 & 1 & 0 \\ -1 & 0 & 0 \\ 0 & 0 & 0 \end{vmatrix} \tag{120}$$

It is easy to find out that, for such a case:
$$L^i_k L^k_j = L_k L_j \delta^{ik} - \delta^i_j = \begin{vmatrix} -1 & 0 & 0 \\ 0 & -1 & 0 \\ 0 & 0 & 0 \end{vmatrix} \tag{121}$$

Hence, for initial isotropic material:
$$\sigma^i_j = -\sigma_S \begin{vmatrix} 1 & 0 & 0 \\ 0 & 1 & 0 \\ 0 & 0 & 0 \end{vmatrix} \tag{122}$$

It is equivalent to yield condition:
$$|\sigma_1 - \sigma_3| = \sigma_S, \text{ or } |\sigma_2 - \sigma_3| = \sigma_S \tag{123}$$

Therefore the yield condition (119) can be generalized as a form similar with classical mechanics:
$$|\sigma^i_j| \leq |\sigma_S L^i_m L^m_j| \tag{124}$$

Recalling the discussion about plasticity and the conclusion about the symmetry of static stress, for a given material, as the parameters $\sigma_S, \lambda, \mu$ are known, the plastic deformation can be calculated by equations (116) and (118).

For general material, the yield stress is not a constant but stress dependent. Metal, rock, and so-called non-linear behavior material are such a kind of continuum. This is easy to understand as the effects of temperature are not taken in above discussion. In this case, as the $S_j^i = \tilde{S}_j^i$ for two forms of Chen's decomposition and both forms must meet the motion equations (68), by combing the two sets of motion equations, after subtracting algebra operation between corresponding equations, it is easy to get that the motion equations for the six rotation parameters and the six components of yield stress are:

$$(\Delta \tilde{\sigma}_j^i)\big|_j = 0 \tag{125-1}$$

$$(\Delta \tilde{\sigma}_j^i)\big|_i = 0 \tag{125-2}$$

$$e_{ijk}(\frac{1}{\cos\theta}\tilde{R}_l^j - R_l^j)\sigma_{Sk}^l = 0 \tag{125-3}$$

$$\sigma_{Sj}^i = E_{jl}^{ik}(1-\cos\Theta_c)L_m^l L_k^m = E_{jl}^{ik}(\frac{1}{\cos\theta}-1)\tilde{L}_m^l \tilde{L}_k^m \tag{125-4}$$

Note that at the cracking point, where:

$$\Delta \tilde{\sigma}_j^i = E_{jl}^{ik}(\frac{1}{\cos\theta}\tilde{R}_k^l - R_k^l) \tag{126}$$

For related tensors, please see equations (3-12).

In fact, equations (125) give the evolution equation for cracking.

**10 Evolution from Cracking to Flow**

After cracking or fatigue, there is local irreversible deformation. The material still forms continuum. However, the deformation from Chen's form-one becomes Chen's form-two.

$$F_j^i = \tilde{S}_j^i + (\cos\theta)^{-1}\tilde{R}_j^i \tag{11}$$

The motion equation becomes:

$$(\sigma_l^i)\big|_l = \frac{\partial}{\partial t}(\rho\frac{\partial U^i}{\partial t}) \tag{127-1}$$

$$(\sigma_j^i)\big|_i = \frac{\partial}{\partial t}[(\rho\frac{\partial U^i}{\partial t})g_{il}^0(\tilde{S}_j^l + \frac{1}{\cos\theta}\tilde{R}_j^l)] \tag{127-2}$$

$$e_{ijk}F_l^j\sigma_k^l = 0 \tag{127-3}$$

where, $\sigma_j^i = E_{jl}^{ik}(F_k^l - \delta_k^l)$.

For a flow problem, the intrinsic stretching strain $\tilde{S}_j^i$ is required to be zero. So, the flow is viewed as pure intrinsic local rotation, here. Physically, it means that the flow material element has no intrinsic stretching, although there still is non-zero displacement, as:

$$\varepsilon_j^i = \frac{1}{2}(u^i\big|_j + u^i\big|_j^T) = (\frac{1}{\cos\theta}-1)(\tilde{L}_k^i \tilde{L}_j^k + \delta_j^i) \tag{128}$$

Comparing with pure rotation in Chen's form-one, this kind of flow should be named as macro-flow, while the pure rotation of Chen's form-one should be named as micro-flow. The difference between them is that the Chen's form-one pure rotation has no gauge variation, but the Chen's form-two pure rotation do have gauge variation.

When the initial co-moving dragging coordinator system is taken to be the standard rectangular coordinator system, the motion equation can be rewritten as:

$$(\sigma_l^i)\big|_l = \frac{\partial}{\partial t}(\rho\frac{\partial U^i}{\partial t}) \tag{127-1}$$

$$(\sigma_j^i)\big|_i = \frac{\partial}{\partial t}(\rho\frac{1}{\cos\theta}\frac{\partial U^i}{\partial t}\tilde{R}_j^i) \tag{127-2}$$

$$e_{ijk}\tilde{R}_l^j\sigma_k^l = 0 \tag{127-3}$$

where, $\sigma_j^i = E_{jl}^{ik}(\frac{1}{\cos\theta}\tilde{R}_k^l - \delta_k^l)$.

Firstly, it should point out that for:

$$\tilde{L}^i_j = \begin{vmatrix} 0 & \tilde{L}_3 & -\tilde{L}_2 \\ -\tilde{L}_3 & 0 & \tilde{L}_1 \\ \tilde{L}_2 & -\tilde{L}_1 & 0 \end{vmatrix} \tag{128}$$

It is easy to find out that, for such a case:

$$\tilde{L}^i_k \tilde{L}^k_j = \tilde{L}_k \tilde{L}_j \delta^{ik} - \delta^i_j \tag{129}$$

Let: $\tilde{L}_i = \tilde{L}^i$, the classical macro flow motion strain becomes:

$$\varepsilon^i_j = \frac{1}{2}(U^i\big|_j + U^i\big|_j^T) = (\frac{1}{\cos\theta} - 1)\tilde{L}^i \tilde{L}_j \tag{130}$$

For the case: $\tilde{L}_3 = 1$, $\tilde{L}_1 = \tilde{L}_2 = 0$, the only non-zero classical strain component is:

$$\varepsilon^3_3 = \frac{1}{2}(U^3\big|_3 + U^3\big|_3^T) = (\frac{1}{\cos\theta} - 1) \tag{131}$$

It means that the continuum is expanding along $x^3$ direction. Here, it will be called intrinsic flow along $x^3$ direction. For such a special macro-flow, the $U^3$ is the only one displacement component. In this case, one has:

$$\tilde{R}^i_j = \begin{vmatrix} \cos\theta & \sin\theta & 0 \\ -\sin\theta & \cos\theta & 0 \\ 0 & 0 & 1 \end{vmatrix} \tag{132}$$

Hence,

$$\tilde{W}^i_j = \frac{1}{\cos\theta}\tilde{R}^i_j - \delta^i_j = \begin{vmatrix} 0 & \tan\theta & 0 \\ -\tan\theta & 0 & 0 \\ 0 & 0 & \frac{1}{\cos\theta} - 1 \end{vmatrix} \tag{133}$$

For initial isotropic simple material, the stress is:

$$\sigma^i_j = (\lambda + 2\mu)(\frac{1}{\cos\theta} - 1)\delta^i_j + 2\mu\begin{vmatrix} 0 & \tan\theta & 0 \\ -\tan\theta & 0 & 0 \\ 0 & 0 & 0 \end{vmatrix} \tag{134}$$

Hence, the motion equation becomes:

$$(\lambda + 2\mu)\frac{\partial}{\partial x^3}(\frac{1}{\cos\theta}) = \frac{\partial}{\partial t}(\rho\frac{\partial U^3}{\partial t}) \tag{135-1}$$

$$(\lambda + 2\mu)\frac{\partial}{\partial x^2}(\frac{1}{\cos\theta}) + 2\mu\frac{\partial(\tan\theta)}{\partial x^1} = 0 \tag{135-2}$$

$$(\lambda + 2\mu)\frac{\partial}{\partial x^1}(\frac{1}{\cos\theta}) - 2\mu\frac{\partial(\tan\theta)}{\partial x^2} = 0 \tag{135-3}$$

$$(\lambda + 2\mu)\frac{\partial}{\partial x^3}(\frac{1}{\cos\theta}) = \frac{\partial}{\partial t}(\rho\frac{1}{\cos\theta}\frac{\partial U^3}{\partial t}) \tag{135-4}$$

The equation (135-1) and (135-4) can be met only when:

$$\rho(\frac{1}{\cos\theta} - 1)\frac{\partial U^3}{\partial t} = D \tag{136}$$

Where, D is a constant. For such a case, if $\rho$ is time-dependent, $\theta$ is time-dependent also. Therefore, when the local rotation angular $\theta$ is constant the material element has constant linear momentum. That means the material will fly away from its original position. This is the common phenomenon during cracking experiments. Once the material goes away, the continuum forms macro-cracks. Then this equation describes the development of cracking tip motion. Recalling that the critical angular is determined by material feature, it is easy to understand this point.

It shows that, for such a kind of flow, the momentum of flow along the local rotation direction and the local rotation angular forms invariant with time. This is a very important result to understand the cracking development phenomenon. The parameter should be an intrinsic feature of material, from physical consideration.

For such a kind of constant flow, that is cracking-tip development, the local rotation angular meets:

$$(\lambda+2\mu)\sin\theta\cdot\frac{\partial\theta}{\partial x^2}+2\mu\frac{\partial\theta}{\partial x^1}=0 \tag{137-1}$$

$$(\lambda+2\mu)\sin\theta\cdot\frac{\partial\theta}{\partial x^1}-2\mu\frac{\partial\theta}{\partial x^2}=0 \tag{137-2}$$

$$(\lambda+2\mu)\frac{\partial}{\partial x^3}(\frac{1}{\cos\theta})=D\frac{\partial}{\partial t}(\frac{\cos\theta}{1-\cos\theta}) \tag{137-3}$$

In fracture mechanics, this flow corresponds to the development of cracking. For a dynamic cracking process, the local rotation angular varies with time and location. Hence, the equation (137-3) will determine the traveling speed of cracking-tip along rotation axe $x^3$ direction. For this purpose, rewritten equation (137-3) as:

$$\frac{\lambda+2\mu}{\cos^2\theta}\frac{\partial\theta}{\partial x^3}=-\frac{D}{(1-\cos\theta)^2}\frac{\partial\theta}{\partial t} \tag{138}$$

It is a highly non-linear wave equation. For a fixed $\theta$, the phase velocity of its perturbation solution is:

$$V_{tip}=-\frac{(\lambda+2\mu)(1-\cos\theta)^2}{D\cdot\cos^2\theta} \tag{139}$$

For the cracking-tip, as D is positive, the angular wave is radiated from the tip. For the far away point from tip, as D is negative, the angular wave travels toward the cracking-tip.

For the tip-point, by equation (136), the velocity can be rewritten as:

$$V_{tip}=-\beta\frac{V_P^2}{V^3} \tag{140-1}$$

where:

$$V_P^2=\frac{\lambda+2\mu}{\rho},\beta=\frac{1-\cos\theta}{\cos\theta},V^3=\frac{\partial U^3}{\partial t} \tag{140-2}$$

and $V_P$ is the P-wave velocity of material. Therefore, the tip wave velocity is determined by intrinsic parameters of material.

For more general case, it is valuable to form an invariant as:

$$V^3\cdot V_{tip}=-\beta V_P^2 \tag{141}$$

This equation well explains the tip wave in fracture mechanics and experiment of cracking development.

The equation (137-1) and (137-2) show that for constant flow the intrinsic rotation angular variants on the normal plane of flow direction. In fluid mechanics, it is called vortex flow. They can be transformed into plane Poisson's equation forms:

$$\frac{\partial^2\theta}{(\partial x^1)^2}+\frac{\partial^2\theta}{(\partial x^2)^2}=0 \tag{142-1}$$

$$\frac{\partial^2\cos\theta}{(\partial x^1)^2}+\frac{\partial^2\cos\theta}{(\partial x^2)^2}=0 \tag{142-2}$$

They show that both $\theta$ and $\cos\theta$ are related with a potential function. Their gradient fields are conservative field. So, only a narrow family of $\theta$ solution function can meets the equation (142), that means the local rotation at cracking-tip point is very deterministic rather than a random process..

As the only one non-zero classical strain component is:

$$\varepsilon_3^3=\frac{1}{2}(U^3\big|_3+U^3\big|_3^T)=(\frac{1}{\cos\theta}-1) \tag{131}$$

By equation (142-2), it is easy to get:

$$\frac{\partial^2}{(\partial x^1)^2}(\frac{1}{1+\varepsilon_3^3})+\frac{\partial^2}{(\partial x^2)^2}(\frac{1}{1+\varepsilon_3^3})=0 \tag{143}$$

For cracking deformation along $x^3$ direction, as $g^{33}=\sqrt{\frac{1}{1+\varepsilon_3^3}}$, it can be rewritten as:

$$\frac{\partial^2}{(\partial x^1)^2}(g^{33})^2+\frac{\partial^2}{(\partial x^2)^2}(g^{33})^2=0 \tag{144}$$

It shows that the square of current area gauge meets Poisson's equation on cracking plane. This phenomenon has well observed in experiment, but not well formulated.

Finally, it is concluded that the cracking-tip behaves as a radiating wave along its local rotation direction and behaves as a conservative filed on the cracking plane.

Many results and conclusions mentioned above are completely new and important progressive. They can be checked by new experiment, say new because the related theoretic results have been compared with the formulated results available usually in common documents.

## 11 Summary and Conclusion

Firstly, the explicit expressions of strain definition in well-known non-linear large deformation theories are obtained by Chen's additive decomposition of displacement gradient. Based on the forms of strain definition, the geometrical aspects between polar-decomposition and additive-decomposition are discussed. After a comparison, the dis-advantage and advantage for different strain definition becomes clear. The research, there after, introduces the Chen's strains which have two forms, one is based on Chen's form-one describing pure elastic deformation, one is based on Chen's form-two describing plastic deformation.

Secondly, it shows that the path-dependency of plastic deformation can be naturally expressed by the stress defined on Chen's form-two strain definition. Therefore, the research goes to develop the motion equation of large deformation. The research finds that the classical static motion equation is correct and the static stress and strain must be symmetric. However, for dynamic deformation, the classical motion equation only meets linear momentum conservation. As the angular momentum conservation is mist in the classical motion equation, the classical dynamic theory of deformation is only an infinitesimal approximation. It is not correct even for infinitesimal deformation if the time duration is long enough that the fatigue appears in material. By this discussion, the value of motion equation gotten in this research is well praised.

Thirdly, based the motion equation, the fatigue problem of material is discussed by dynamic motion equation. The contradiction between static symmetry and dynamic non-symmetry makes the existence of fatigue. The research explicitly formulates the fatigue development equations for two typical cases: fatigue caused by pure stretching deformation action, and fatigue caused by pure rotation deformation action. These results can be directly checked by experiments and observation. Hence, it is testable.

Generally, the deformation of continuum will be initially elastic, so it follows Chen's form-one. No matter what kind of cause, when the local rotation angular reaches a critical value, which is an intrinsic parameter of material, the deformation has two decomposition forms. This feature is used to discuss the cracking condition equation. It shows that the yield fracture conditions of different forms in classical mechanics can be naturally explained by the cracking condition equation gotten by this research. More important is that the research gives the cracking condition equation in closed-form, so the fracture problem can be solved exactly by cracking deformation equations. These results can be directly checked by experiments and observation.

For the damaged or cracked continuum, the deformation is completely described Chen's form-two. The research develops the related cracking-tip motion equations. Without lose of generality, the research studies a special case where one coordinator is taken as the axe of local rotation direction. The results show that the local rotation angular and its cosine-function meet Poisson's equation on cracking plane. It also shows that the square of area gauge along the rotation direction meets Poisson's equation on the cracking plane. The research shows that the cracking-tip motion equation along the rotation axe is a highly non-linear wave equation. Its perturbation wave speed is obtained. It is founded that the tip wave speed is completely determined by material feature and the linear momentum at cracking-tip. In fact, the research shows that the local rotation angular and the cracking-tip linear momentum form an invariant for a given material. The displacement component along the rotation axe behaves as a flow. These results can be directly checked by experiments and observation.

Therefore, the research expresses the mechanics of continuum deformation evolution from elastic deformation, through fatigue or plastic deformation, after cracking, to cracking-tip development, and finally does to flow motion. It forms a unified finite deformation mechanics theory for wide range of deformation type. Not only many old problems on theory be solved, but also many new results, ready applicable for fatigue, fracture, and cracking development, be gotten. It is believed that the new theoretic results will promote our research on the non-classical deformation, especially the so-called non-linear mechanics of continuum.

**References**
[1]. Truesdell C. The main problem in the finite theory of elasticity, ed. Truesdell C., *Foundations of elasticity theory,* Gorden & Breach Science Pub., 1966
[2]. Casey J, Naghdi P M. A remark on the use of the decomposition F=$F_e F_p$ in plasticity, *Journal of Applied Mechanics* **47**: 672-675, 1980
[3]. Lubliner J., *Plasticity Theory, New* York: Macmillan Pub. Comp. 1990
[4]. Eringin A C, Kafadar C B., Polar field theory, ed Eringin A C, *Continuum Physics IV*, New York: Academic Press., 2-75, 1976
[5]. Steinmann P., A micropolar theory of finite deformation and finite rotation multiplicative elastoplastigity. *Intern. J. Non-Linear Mechanics* **32**: 103-119, 1994
[6]. Grammenoudis P, Tsakmakis C., Hardening rules for finite deformation micropolar plasticity: Restrictions imposed by the second law of thermodynamics and the postulate of Ii'iushin, *Continuum Mech. Thermodyn.* **13**: 325-363, 2001
[7]. Chen Zhida. Geometric Theory of Finite Deformation Mechanics for Continuum. *Mechanica Sinica,* ,No.2, 107-117, , (In Chinese), 1979
[8]. Chen Zhida, Limit Rotation Expression in Non-linear Field Theory of Continuum. *Applied Mathematics and Mechanics*, 1996 No.7, 959-968, (In Chinese), 1986
[9]. Chen Zhida. *Rational Mechanics—Non-linear Mechanics of Continuum*. Xizhou: China University of Mining & Technology Publication, (In Chinese) 1987
[10]. Chen Zhida. *Rational Mechanics.* Chongqin: Chongqin Publication, (In Chinese) 2000
[11]. Xiao Jianhua, Decomposition of displacement gradient and strain definition, *Advance in Rheology and its Application(2005)*, 864-868, Science Press USA Inc.,2005
[12]. Guo Zhonghen, Lian Haoyin. *Non-compatible Theory of Deformation*. Chongqin: Chongqin Publication, 1989
[13]. Milklowitz, J., *Elastic waves and waveguides*, North-Holland Pub. Comp., 1978
[14]. Einstein, A., Relativistic Theory of Non-Symmetric Field, *Einstein's Work Collection* (Vol.2), Commercial Pub., p565 (in Chinese) , 1977
(Original: Einstein, A., *Meaning of Relativity*, (5$^{th}$ edition), p133-166, 1954)
[15]. Truesdell C., *The mechanical foundations of elasticity and fluid dynamics,* Gorden & Breach Science Pub., 1966
[16]. Dubrovin, B. A., A. T. Fomenko, & S. P. Novikov, *Morden Geometry – Methods and application. Part I: The geometry of surfaces, transformation groups and fields*. New York: Springer-Verlag, p259, 1984
[17]. Diepolder W, Mannl V, Lippmann H. The Cosserat continuum, a modle for grain rotations in metals?. *International Journal of Plasticity (1991)***7**, 313-328, 1991
[18]. Biot A M. *Mechanics of Incremental Deformations*. New York :John Wiley & Sons Inc.,1965
[19]. Kuang Zhengban, *Non-linear Continuum Mechanics,* Xi'an: Xi'an Jiaotong University Pub. (In Chinese) 1989
[20]. Wu Jike, Su Xianyue. *Stability of Elastic System,* Beijing: Science Publication, (In Chinese), 1994
[21]. Truesdell C. General and exact theory of waves in finite elastic strain. In Truesdell C ed. *Problems of non-linear elasticity,* Gordon & Breach Science Pub.,1965
[22]. Crampin S & J. H. Lovell. A decade of shear-wave splitting in the Earth's crust: What does it mean? What use can we make it? And what should we do next? *Geophy. J. Int.*, **107:** 387～407, 1991
[23]. Toupin R Aand B. Bernstein. Sound Waves in Deformed Perfectly Elastic Materials. In Truesdell C. ed. *Foundations of Elasticity Theory.* Gorden & Breach Science Pub. 1966
[24]. Crampin S. A review of wave motion in anisotropic and cracked elastic-media, *Wave Motion,***3**:343～391, 1981
[25]. Lou Zhiwen, *Introduction to Damage Mechanics of Material,* Xi'an: Xi'an: Xi'an Jiaotong University Pub. (In Chinese) 1991

**Appendix A    Mathematic Feature of Deformation Tensor**

For continuum, each material point can be parameterized with continuous coordinators $x^i, i=1,2,3$. When the coordinators are fixed for each material point no matter what motion or deformation happens, the covariant gauge field $g_{ij}$ at time $t$ will define the configuration in the time. The continuous coordinators endowed with the gauge field tensor define a co-moving dragging coordinator system.

The initial configuration gauge $g_{ij}^0$ defines a distance geometric invariant:

$$ds_0^2 = g_{ij}^0 dx^i dx^j \tag{A-1}$$

The symmetry and positive feature of gauge tensor insures that there exist three initial covariant base vectors $\vec{g}_i^0$ make:

$$g_{ij}^0 = \vec{g}_i^0 \cdot \vec{g}_j^0 \tag{A-2}$$

For current configuration, three current covariant base vectors $\vec{g}_i$ exist which make:

$$g_{ij} = \vec{g}_i \cdot \vec{g}_j \tag{A-3}$$

The current distance geometric invariant is:

$$ds^2 = g_{ij} dx^i dx^j \tag{A-4}$$

For each material point, there exists a local transformation $F_j^i$, which relates the current covariant base vectors with initial covariant base vectors:

$$\vec{g}_i = F_i^j \vec{g}_j^0 \tag{A-5}$$

So, the current covariant gauge tensor can be expressed as:

$$g_{ij} = F_i^k F_j^l g_{kl}^0 \tag{A-6}$$

In Riemann geometry, contra-variant gauge tensor $g^{ij}$, $g^{0ij}$ can be introduced, which meets condition:

$$g^{il} g_{jl} = \delta_j^i, \quad g^{0il} g_{jl}^0 = \delta_j^i \tag{A-7}$$

Similarly, contra-variant base vectors $\vec{g}^i$, $\vec{g}^{0i}$ can be introduced for current configuration and initial configuration respectively. Mathematically, there are:

$$g^{ij} = \vec{g}^i \cdot \vec{g}^j, \quad g^{0ij} = \vec{g}^{0i} \cdot \vec{g}^{0j} \tag{A-8}$$

There exists a local transformation $G_j^i$, which relates the contra-variant base vectors:

$$\vec{g}^i = G_j^i \vec{g}^{0j} \tag{A-9}$$

So, the current contra-variant gauge tensor can be expressed as:

$$g^{ij} = G_k^i G_l^j g^{0kl} \tag{A-10}$$

By equations (A-6), (A-7), and (A-10), it is easy to find out that:

$$G_l^i F_j^l = \delta_j^i \tag{A-11}$$

Hence, the transformation $F_j^i$ relates the initial contra-variant base vectors with current contra-variant base vectors in such a way that:

$$\vec{g}^{0i} = F_j^i \vec{g}^j \tag{A-12}$$

Therefore, the transformation $F_j^i$ is a mixture tensor. Its lower index represents covariant component in $g_{ij}^0$ configuration, its upper index represents contra-variant component in $g_{ij}^0$ configuration.

Similar discussion shows that the local transformation $G_j^i$ is a mixture tensor, lower index represents covariant component in $g_{ij}$ configuration, upper index represents contra-variant component in $g_{ij}$ configuration. It is easy to find that:

$$\vec{g}_i^0 = G_i^j \vec{g}_j \tag{A-13}$$

Other two important equations are:

$$F_j^i = \vec{g}^{0i} \cdot \vec{g}_j \tag{A-14}$$

$$G_j^i = \vec{g}^i \cdot \vec{g}_j^0 \tag{A-15}$$

Bt these equations, $F_j^i$ can be explained as the extended-Kronecker-delta in that it's the dot product of contra-variant base vector in initial configuration and covariant base vector in current configuration. When the two configurations are the same, it becomes the standard Kronecker-delta. For the $G_j^i$, similar interpretation can be made.

From these equations to see, geometrically, it is found that the $F_j^i$ measures the current covariant base vector with the initial contra-variant base vector as reference. When stress tensor is defined as:

$$\sigma_j^i = E_{jl}^{ik}(F_k^l - \delta_k^l) \tag{A-16}$$

Its mechanic meaning can be explained as the lower index represents component of surface force in current direction, and the upper index represents the initial surface normal which surface force acts on. It takes the initial surface as surface force reference.

The $G_j^i$ measures the initial covariant base vector with the current contra-variant base vector as reference. The corresponding stress can be explained as the lower index represents component of surface force in initial direction, and the upper index represents the current surface normal which surface force acts on.

For such a kind of mixture tensor, it is defined on the same point with different configurations (that is initial and current). So, the name of two-point tensor given by C Truessdell is not correct. This mis-interpretation has caused many doubts cast on the feature of transformation tensor. Some mathematician even said that the mixture tensor is meaningless.

However, during treating the non-symmetric field theory, Einstein believes that the use of mixture tensor is more reasonable that the pure covariant or pure contra-variant tensor. Recently, the concept of bi-tensor is introduced in physics topic. So, we have sound reason to use mixture tensor in continuum mechanics, as it can give clear physical meaning for the definition of stress.

Note that the transformation $F_j^i$ is completely determined by the deformation measured in initial configuration with gauge tensor $g_{ij}^0$. Mathematically, the covariant differentiation is taken in the initial geometry also, although the physical meaning of $F_j^i$ is that it relates two configurations.

It is valuable to point out that, if the initial coordinator system is taken as Cartesian system, the $F_j^i$ can be transformed into pure covariant form:

$$F_{ij} = \vec{g}_i^0 \cdot \vec{g}_j \tag{A-17}$$

The $G_j^i$ can be transformed into pure contra-variant form:

$$G^{ij} = \vec{g}^i \cdot \vec{g}^{0j} \tag{A-18}$$

Although it is acceptable in form for the special case of taking Cartesian system as the initial coordinator system, the intrinsic meaning of deformation tensor is completely destroyed by such a formulation. That may be the main reason for the rational mechanics constructed by C Truessdell et al in 1960s.

Mathematically, once the initial gauge field is selected, the current gauge field is to be obtained by the given physical deformation. In this sense, the current gauge field is viewed as the physical field. So, the covariant differentiation is taken respect with the initial configuration.

Truessdell argued that the covariant differentiation should be taken one index in initial configuration, another index in current configuration. This concept had strongly effects on the development of infinite deformation mechanics. Such a kind of misunderstanding indeed is caused by the equations (A17-18).

Historically, Chen Zhida is the first one to clear the ambiguity caused by equations (A-17) and (A-18) systematically. His monograph "Rational Mechanics"(1987) treats this topic in depth.

There are many critics about the mathematics used in Chen's rational mechanics theory. The most comment one is that: for a coordinator transformation

$$dx^i = A_j^i dX^j, \quad dX^i = (A_j^i)^{-1} dx^j \tag{A-19}$$

the covariant and contra-variant components of tensor is defined by its transformation from new coordinator system to old coordinator system follows $A_j^i$ or $(A_j^i)^{-1}$ formulations. However, such a tensor definition is to make:

$$ds^2 = g_{ij} dx^i dx^j = G_{ij} dX^i dX^j \tag{A-20}$$

be invariant. Such a tensor describes a continuum without any motion. In fact, such a tensor feature is only to say the object indifference for coordinator system selection.

However, many physicists and mechanists treat motion of continuum as the transformation $A_j^i$ or

$(A^i_j)^{-1}$. Mathematically, the equation (A-20) can be rewritten as:

$$ds^2 = g_{ij}dx^i dx^j = g_{kl}A^k_i A^l_j dX^i dX^j = G_{ij}dX^i dX^j \tag{A-21}$$

or:

$$ds^2 = G_{ij}dX^i dX^j = G_{kl}(A^k_i)^{-1}(A^l_j)^{-1}dx^i dx^j = g_{ij}dx^i dx^j \tag{A-22}$$

It is clear that the covariant invariant feature must be maintained by the coordinator system choice. It has no any meaning of motion. In Chen's geometry, for time parameter $t$,

$$ds^2(t) = g_{ij}(t)dx^i dx^j = g_{kl}(0)F^k_i(t)F^l_j(t)dx^i dx^j \tag{A-23}$$

The gauge field is time dependent, while the coordinator is fixed (called intrinsic coordinator).

The equation (A-23) can be rewritten as:

$$ds^2(t) = g_{ij}(t)dx^i dx^j = g_{kl}(0)F^k_i(t)F^l_j(t)dx^i dx^j = g_{kl}(0)dX^k(t)dX^l(t) \tag{A-24}$$

It is similar in form with equation (A-21). But their mechanical interpretation is strikingly different.

**Appendix B Relationship between Chen's Constitutive and Classical Constitutive Equations**

Here, for simplicity, the material is idea isotropic simple elastic material and the initial coordinator system is standard rectangular coordinator system. For the initial reference configuration, Chen's constitutive equation is:

$$\sigma^i_j = E^{ik}_{jl}(F^l_k - \delta^l_k) \tag{B-1}$$

It defines two stresses for pure elastic deformation:

$$\tilde{\sigma}^i_j = E^{ik}_{jl} S^l_k \tag{B-2}$$

$$\hat{\sigma}^i_j = E^{ik}_{jl}(R^l_k - \delta^l_k) \tag{B-3}$$

For pure elasticity, classical stress is:

$$^C\sigma_{ij} = E^{ik}_{jl} \frac{1}{2}(U^l|_k + U^l|_k^T) = \tilde{\sigma}^i_j + (1-\cos\Theta)E^{ik}_{jl}L^m_k L^l_m \tag{B-4}$$

When the local rotation is zero, there is no difference. For large deformation, classical theory introduces additional elasticity. Chen's theory shows that large deformation will still be the same elasticity parameter. Physically, the large deformation, if the temperature is constant, will not change the intrinsic feature of material. So, Chen's theory is much more soundness.

Another difference is for the elasticity constants. In Chen's theory, the elasticity constant tensor is a mixture tensor, its isotropic form is:

$$E^{ik}_{jl} = \lambda \delta^i_j \delta^k_l + 2\mu \delta^i_l \delta^k_j \tag{B-5}$$

Where, in classical mechanics, the elasticity constant tensor is defined as:

$$C_{ijkl} = \lambda \delta_{ij}\delta_{kl} + \mu\delta_{ik}\delta_{jl} + \mu\delta_{il}\delta_{jk} \tag{B-6}$$

Therefore, the classical form will not accept non-symmetry stress. This problem is overcome in elastic wave theory to define:

$$C_{ijkl} = \lambda \delta_{ij}\delta_{kl} + 2\mu\delta_{ik}\delta_{jl} \tag{B-7}$$

Its function is same as (B-5) form, but is not so geometrically soundness, because for pure covariant tensor the form (B-6) indeed is the unique isotropic form.

As this definition suits not only for elastic wave, but also suits for explosive waves, it is reasonable to say the form (B-5) has no contradiction with observation. Further more, it shows that the form (B-6) is not physically correct. The problem is that it should be not a pure covariant tensor.

Truessdell, may viewing this problem, gave a form:

$$C_{kl}^{ij} = \lambda \delta^{ij} \delta_{kl} + \mu \delta_k^i \delta_l^j + \mu \delta_l^i \delta_k^j \tag{B-8}$$

Therefore, Truessdell strongly against the introduction of non-symmetry elasticity stress.

As the non-symmetry stress does function in infinitesimal S-wave motion, so the forms (B-6) and (B-8) must be abandon for dynamic deformation motion.

Therefore, physically and geometrically, the form (B-5) is the unique form for isotropic elasticity.